\pdfoutput=1

\documentclass[11pt]{article}

\usepackage[]{ACL2023}

\usepackage{times}
\usepackage{latexsym}

\usepackage[T1]{fontenc}

\usepackage[utf8]{inputenc}

\usepackage{microtype}

\usepackage{inconsolata}

\usepackage{graphicx}
\usepackage{amsmath}
\usepackage{hyperref}

\title{AstroLLaMA \protect\includegraphics[height=2em]{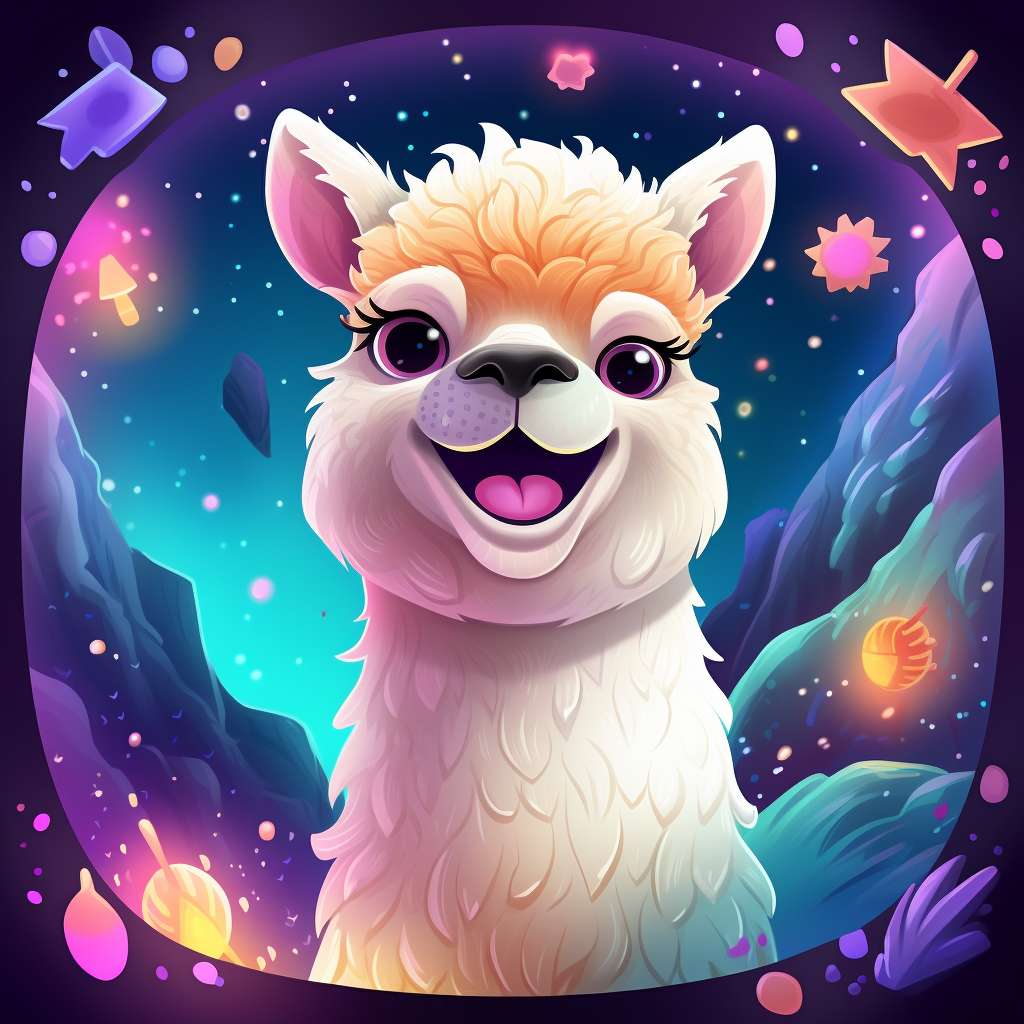}: Towards Specialized Foundation Models in Astronomy}

\newcommand{\customfootnote}[1]{\let\thefootnote\relax\footnote{#1}\let\thefootnote\svthefootnote}

\author{
    Tuan Dung Nguyen\textsuperscript{1, 2*}\protect\customfootnote{Email:joshtn@seas.upenn.edu}, 
    Yuan-Sen Ting\textsuperscript{2, 3*},
    Ioana Ciuc\u{a}\textsuperscript{2*} \\
    \textbf{Charles O'Neill\textsuperscript{2†}}, 
    \textbf{Ze-Chang Sun\textsuperscript{4†}}, 
    \textbf{Maja Jab\l{o}\'{n}ska\textsuperscript{2†}},
    \textbf{Sandor Kruk\textsuperscript{5†}} \\
    \textbf{Ernest Perkowski\textsuperscript{5}}, 
    \textbf{Jack Miller\textsuperscript{2}}, 
    \textbf{Jason Li\textsuperscript{6}}, 
    \textbf{Josh Peek\textsuperscript{7}} 
    \textbf{Kartheik Iyer\textsuperscript{8}}, \\
    \textbf{Tomasz R\'o\.{z}a\'{n}ski\textsuperscript{2,9}}, 
    \textbf{Pranav Khetarpal\textsuperscript{10}},  
    \textbf{Sharaf Zaman\textsuperscript{2}},
    \textbf{David Brodrick\textsuperscript{2}} \\
    \textbf{Sergio J. Rodr\'{i}guez M\'{e}ndez\textsuperscript{2}},
    \textbf{Thang Bui\textsuperscript{2}}, 
    \textbf{Alyssa Goodman\textsuperscript{11}}, 
    \textbf{Alberto Accomazzi\textsuperscript{12}}, \\
    \textbf{Jill Naiman\textsuperscript{13}},
    \textbf{Jesse Cranney\textsuperscript{2}}, 
    \textbf{Kevin Schawinski\textsuperscript{14}},
    \textbf{UniverseTBD} \\
    \textsuperscript{1}University of Pennsylvania, United States \quad 
    \textsuperscript{2}Australian National University, Australia
    \\
    \textsuperscript{3}Ohio State University, United States
    \quad
    \textsuperscript{4}Tsinghua University, China \\
    \textsuperscript{5}European Space Astronomy Centre, Spain
    \quad \\
    \textsuperscript{6}Learning Machines, Australia
    \quad
    \textsuperscript{7}Space Telescope Science Institute, United States
    \\
    \textsuperscript{8}Columbia University, United States 
    \quad
    \textsuperscript{9}Wroc\l{}aw University, Poland
    \\
    \textsuperscript{10}Indian Institute of Technology Delhi, India
    \quad
    \textsuperscript{11}Harvard University, United States
    \\
    \textsuperscript{12}NASA Astrophysics Data System, Harvard \& Smithsonian, United States
    \\
     \textsuperscript{13}University of Illinois at Urbana-Champaign
    \quad
     \textsuperscript{14}Modulos AG}
    
\begin{document}
\maketitle
\renewcommand*{\thefootnote}{*}
\footnotetext{Lead contribution. Email: \href{mailto:joshtn@seas.upenn.edu }{joshtn@seas.upenn.edu }}
\renewcommand*{\thefootnote}{†}
\footnotetext{Major contribution.}

\begin{abstract}
Large language models excel in many human-language tasks but often falter in highly specialized domains like scholarly astronomy. To bridge this gap, we introduce AstroLLaMA, a 7-billion-parameter model fine-tuned from LLaMA-2 using over 300,000 astronomy abstracts from arXiv. Optimized for traditional causal language modeling, AstroLLaMA achieves a 30\% lower perplexity than Llama-2, showing marked domain adaptation. Our model generates more insightful and scientifically relevant text completions and embedding extraction than state-of-the-arts foundation models despite having significantly fewer parameters. AstroLLaMA serves as a robust, domain-specific model with broad fine-tuning potential. Its public release aims to spur astronomy-focused research, including automatic paper summarization and conversational agent development.
\end{abstract}
\section{Introduction}
\label{sec:intro}

The advent of Large Language Models (LLMs) has sparked interdisciplinary interest thanks to a confluence of factors: accumulation of massive datasets, leaps in computational power and breakthroughs in neural architectures. Flagship models like GPT-4 \citep{openaiGPT4TechnicalReport2023}, PaLM \citep{chowdheryPaLMScalingLanguage2022,GoogleAIPaLM} and LLaMA \citep{touvronLLaMAOpenEfficient2023,metaLLaMAOpenFoundation} have exhibited exceptional versatility in a variety of tasks from logical reasoning and comprehension to creative writing, often accomplished via methods like prompting, fine-tuning, and human-in-the-loop reinforcement learning.

The astronomy discipline presents both a unique challenge and a fertile ground for the application of LLMs. First, the corpus of scholarly texts in astronomy likely constitutes but a minuscule portion of the data on which generic LLMs are trained, resulting in limitations like hallucinations in favor of more ``generic'' responses. Second, the nature of astronomical research often involves cross-disciplinary insights due to universally applicable physical processes. When well-curated, LLMs could meaningfully assist in hypothesis generation. 

Existing scales based on in-context prompting and instruction learning, primarily involving GPT-4, have already demonstrated significant potential for generating substantive hypotheses \citep{Ciuca2023a,Ciuca2023b}. Further, the astronomy community's ``open sky'' policy, which grants public access to the majority of its datasets either immediately or after a brief proprietary period \citep{Almeida2023,Fabricius2021}, pairs well with the wealth of resources available in archives like NASA's Astrophysics Data System \citep{Accomazzi2015,Borgman2021}. Such an open-access policy can facilitate deep engagement with the astronomical literature.

Despite their general capabilities, LLMs frequently lag behind specialized, smaller models in domain-specific applications. This disparity stems from two primary factors: (i) the eclectic nature of the training datasets, which dilutes the focus on specialized subjects, and (ii) the design ethos of LLMs as ``foundation models'' meant for subsequent fine-tuning tailored to specific tasks. The existing landscape for fine-tuned LLMs in astronomy remains limited, however. To our knowledge, the only existing specialized model is astroBERT \citep{grezesBuildingAstroBERTLanguage2021}, which has 110 million parameters, trained on nearly 400,000 ADS papers. But as an non-generative model, the utility of astroBERT remains limited to discriminative tasks.

Motivated by these gaps, we present AstroLLaMA, a state-of-the-art generative language model fine-tuned from LLaMA-2. Our model leverages a corpus of 300,000 astronomy abstracts from arXiv and boasts an architecture approximately 67 times larger than that of astroBERT. AstroLLaMA aspires to build upon astroBERT's foundation by offering improved performance in generating specialized information.

\begin{figure}
	\centering
	\includegraphics[width=1\linewidth]{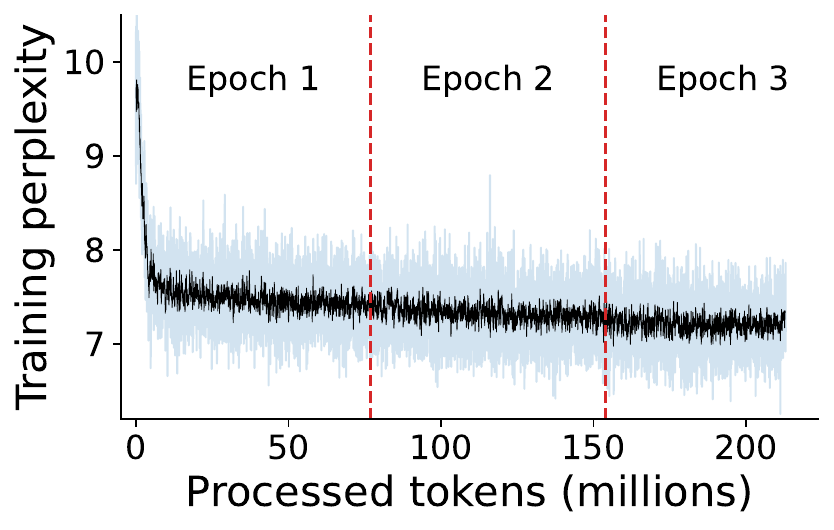}
	\caption{Learning curve of AstroLLaMA during its fine-tuning on the arXiv astrophysics dataset. The Fig.tracks the evolution of perplexity, a measure of the model's next-token prediction performance. The light blue curve shows the training perplexity at each AdamW update step, while the dark black curve provides a smoothed average taken over 10-step intervals.}
	\label{fig:train_ppl}
\end{figure}

\section{AstroLLaMA}

\label{sec:astroLLaMA}

In this section, we discuss AstroLLaMA's implementation, focusing on the curation of its dataset, base model architecture, and fine-tuning settings.

\subsection{Dataset}
\label{sec:astroLLaMA:dataset}

We derive our dataset from the arXiv repository, available on Kaggle.\footnote{\url{https://www.kaggle.com/Cornell-University/arxiv}} Our curated subset focuses on papers classified under the astrophysics category (\texttt{astro-ph}), resulting in a collection of 326,238 articles spanning from April 1992 to July 2023. We extract the these papers' abstracts to form a corpus consisting of approximately 95 million tokens. The median length of these abstracts is 291 tokens. To enable effective model evaluation, we randomly designate 20\% of this curated dataset for testing.

\subsection{Base Model}
\label{sec:astroLLaMA:basemodel}

Our base model is LLaMA-2, a 6.7 billion-parameter model developed by Meta \citep{metaLLaMAOpenFoundation}. Originally trained on a corpus containing 2 trillion tokens, LLaMA-2 features a context window of 4,096 tokens. For tokenization, the model employs a bytepair encoding strategy \citep{sennrichNeuralMachineTranslation2016,kudoSentencePieceSimpleLanguage2018}, incorporating a vocabulary set of 32,000 unique tokens.

\subsection{Fine-tuning Settings}
\label{sec:astroLLaMA:finetuning}

For the fine-tuning phase, we rely on our curated training set described in Section \ref{sec:astroLLaMA:dataset}, which includes 77 million tokens. Special \texttt{[BOS]} (Beginning Of Sequence) and \texttt{[EOS]} (End Of Sequence) tokens are prepended and appended to each training sequence. These sequences are then concatenated and divided into fixed-length chunks, each comprising 512 tokens. 

The fine-tuning process follows the causal language modeling objective employed during the model's pre-training phase. We use the AdamW optimizer \citep{loshchilovDecoupledWeightDecay2018} with hyperparameters $\beta_1 = 0.9, \beta_2 = 0.95, \epsilon = 10^{-5}$ and a batch size of 32. The learning rate follows a cosine schedule with a linear warmup to a peak value of $3 \times 10^{-4}$ in the first 10\% of the optimization steps and a final learning rate of 10\% of its peak. Additional settings include weight decay and gradient clipping values of 0.1 and 1.0, respectively.

We fine-tune LLaMA over nearly three epochs, corresponding to about 230 million processed tokens, using four NVIDIA A100 GPUs, each equipped with 40GB of VRAM. To maximize resource efficiency, we employ 4-bit quantization and utilize LoRA, a technique based on low-rank matrix decomposition \citep{huLoRALowRankAdaptation2021}. We set LoRA's hyperparameters $\alpha$ and dropout rate to 32 and 0.05, respectively. The entire process is facilitated through the Hugging Face Python library.

\begin{figure*}
	\centering
	\includegraphics[width=1\linewidth]{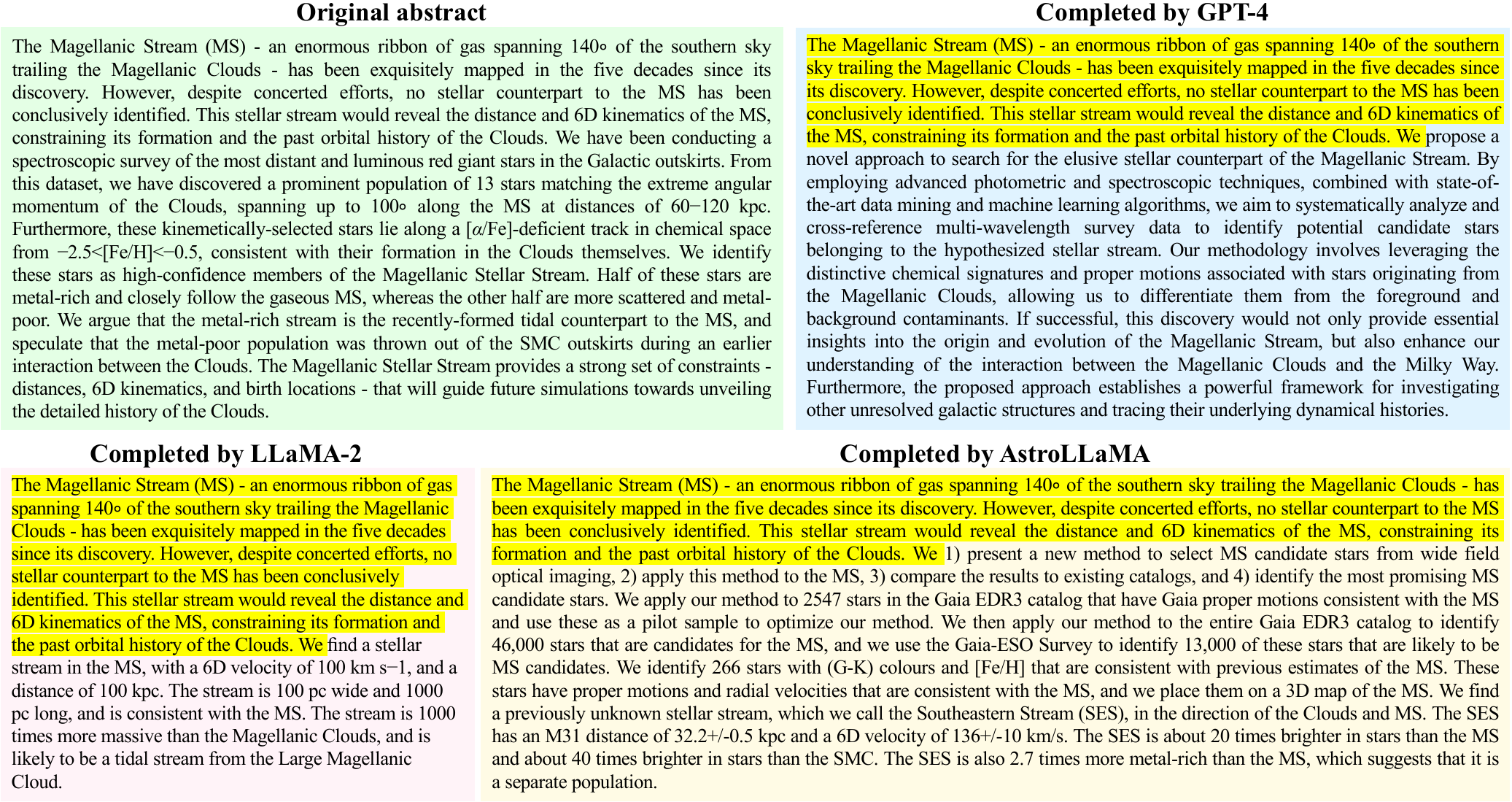}
	\caption{Completion of an abstract from the arXiv database (ID: 2306.15719) using three different models: GPT-4, LLaMA-2, and AstroLLaMA. Each model is prompted with the same short text snippet, highlighted in their respective boxes. GPT-4 tends to produce more generic statements, lacking domain-specific nuance. AstroLLaMA demonstrates the most robust completion, offering more relevant concepts and deeper insights specific to the field of astronomy, thus significantly outperforming LLaMA-2 and GPT-4.}
	\label{fig:generation_prelim}
\end{figure*}

\subsection{Fine-Tuning Evaluation}
\label{sec:astroLLaMA:eval}

Fig. \ref{fig:train_ppl} depicts the performance of AstroLLaMA during its fine-tuning phase. Here, we present perplexity, a commonly used metric for evaluating causal language models. Perplexity is defined as the exponentiation of the training loss, with lower values indicating a better fit.

Our initial observations reveal that LLaMA-2 performs suboptimally on our dataset, with an average perplexity close to 10. By the conclusion of three epoch, AstroLLaMA achieves an average perplexity of 6.55. This represents a 32.5\% reduction in perplexity compared to the base LLaMA-2 model, signifying a substantial improvement in the model's predictive accuracy.

\section{Results}
\label{sec:prelim_results}

As illustrated in the previous section, AstroLLaMA outperforms its non-fine-tuned counterpart, LLaMA-2, in terms of context-awareness during token prediction within astronomy abstracts. To delve deeper into the advantages of fine-tuning, we assess AstroLLaMA's general abilities in two key aspects: \textbf{text generation} and \textbf{embedding space quality}. We compare its performance against multiple models, including LLaMA-2, GPT-4 and GPT-3 (ada-002) to provide a comprehensive evaluation.

Regarding \textbf{text generation}, we task AstroLLaMA, LLaMA-2 and GPT-4 with completing various astronomy-related abstracts, an example of which is presented in Fig. \ref{fig:generation_prelim}. Each model is given the first few sentences of an abstract as a prompt, allowing us to gauge its ability to comprehend the context and generate a meaningful continuation. For GPT-4, we utilize ChatGPT and specifically prompt it to limit the completion to a single paragraph. AstroLLaMA and LLaMA-2 are deployed using standard sampling methods, with the temperature set to 0.3 and a maximum new tokens limit of 1,024. We find that altering the temperature setting does not substantively improve LLaMA-2's results.

Our observations largely echo the patterns depicted in Fig. \ref{fig:generation_prelim}. LLaMA-2 often deviates from the intended context after generating only a short and often off-topic continuation, resulting in inferior completions. While GPT-4 produces more coherent text, its responses are too generic to capture the nuanced understanding required in the astronomy domain. Even when explicitly prompted to focus on astronomy-related topics, GPT-4's generated text remains largely off-target or generically applicable rather than domain-specific.

In stark contrast, AstroLLaMA exhibits remarkable context-awareness in its completions by showing a deep understanding of astronomical concepts. For example, in Fig. \ref{fig:generation_prelim}, AstroLLaMA comprehends that an effective search for stars in the Magellanic Stream involves a three-step process: initial wide-field imaging, followed by refinement using astrometric data from Gaia, and then further curation with spectroscopic data. The model also understands Gaia-ESO is surveying the southern sky and hence can observe (part of) the Magellanic Stream. It also demonstrates nuanced knowledge of the Magellanic Stream, understanding the importance of bifurcation within the stream. As a result, it appropriately completes the text by discussing the southeast stream and exploring metallicity differences to ascertain their origins.

Regarding \textbf{embedding space quality}, we assess models' ability to reflect semantic similarities among astronomy texts. We randomly choose 10,000 abstracts from our dataset and embed them using AstroLLaMA and GPT-3. Specifically, we use OpenAI's API to invoke the text embedding function for GPT-3 (ada-002). To get text embeddings from AstroLLaMA, we pass an input through the model and extract its final hidden states, which contain embeddings for all tokens in the input. Then, we omit the \texttt{[BOS]} token and take the average of all other tokens’ embeddings to get the final result. Finally, for each pair of abstracts we calculate their cosine similarity (the normalised dot product) between on their vector embeddings.

\begin{figure}
	\centering
	\includegraphics[width=1\linewidth]{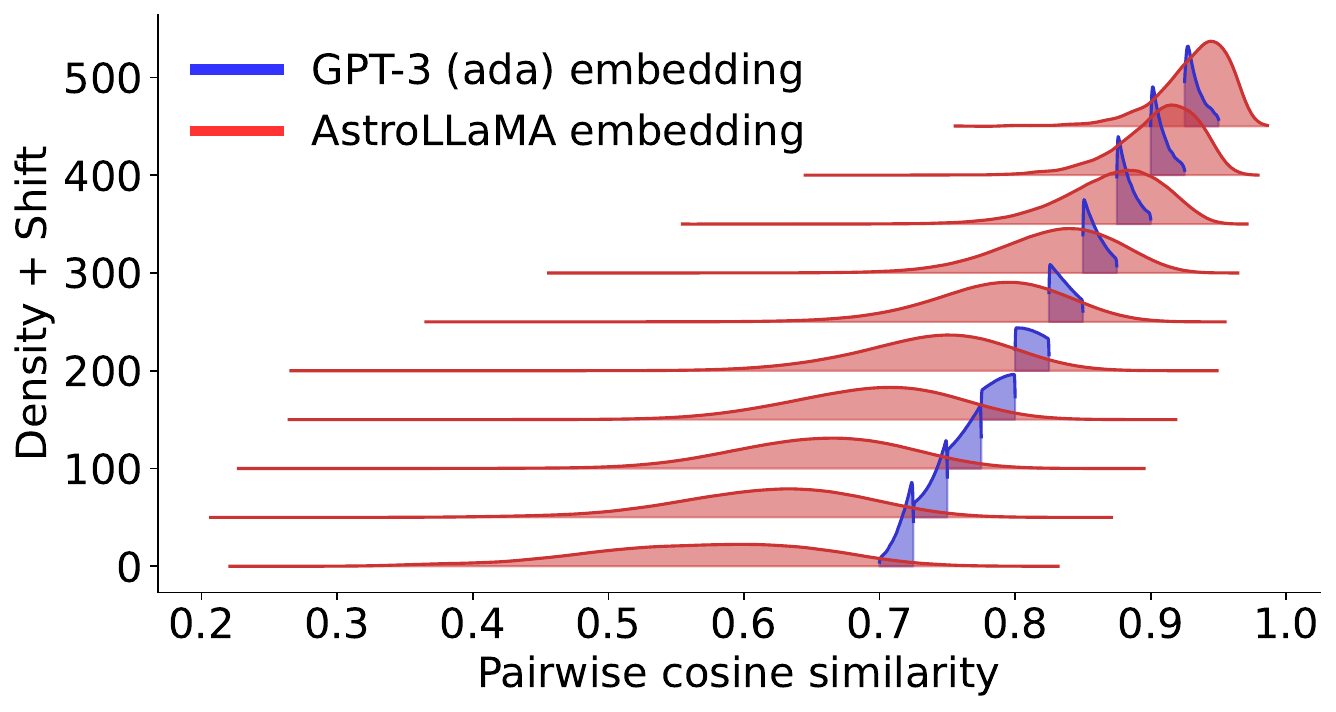}
    \includegraphics[width=1\linewidth]{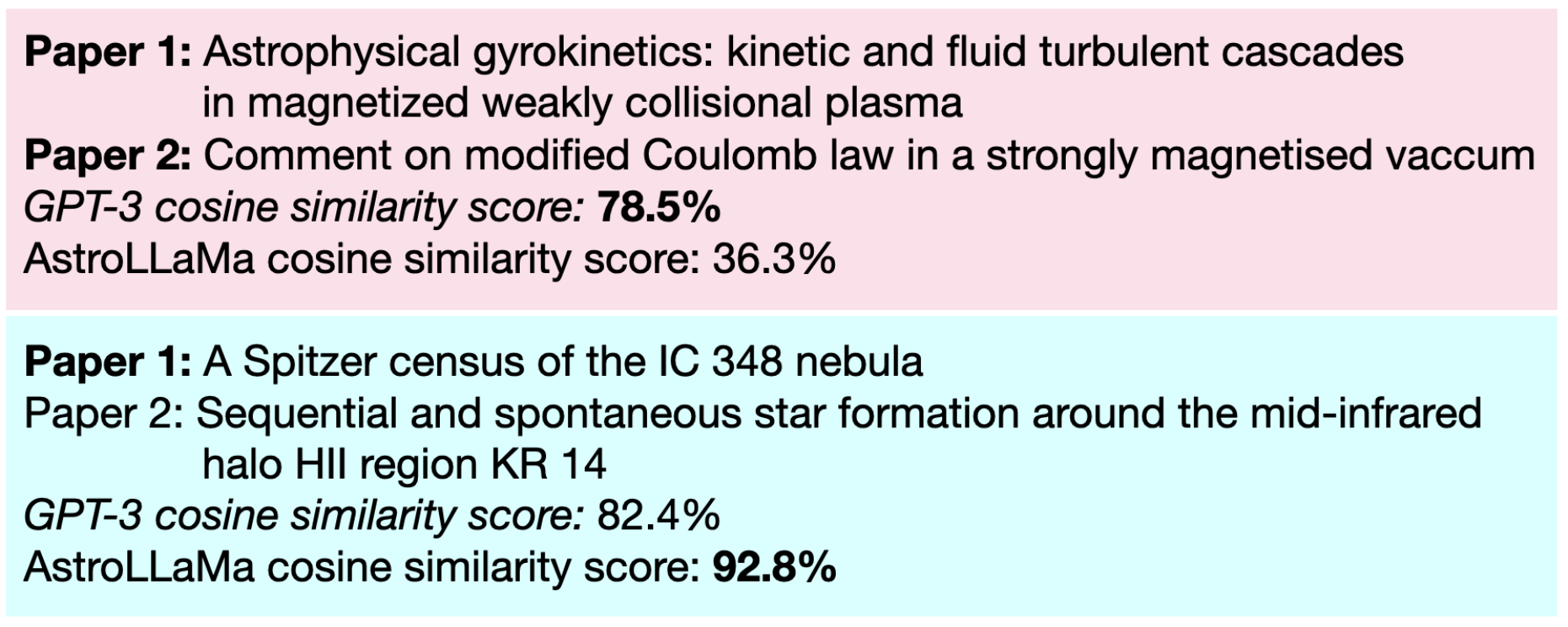}
	\caption{{\it Top:} Distribution of pairwise cosine similarities among 10,000 randomly selected abstracts from our corpus, divided into 10 equal bins based on similarity levels from GPT-3. {\it Bottom:} Two representative examples illustrating divergent cosine similarity values when comparing AstroLLaMA and GPT-3 embeddings.}
	\label{fig:embedding}
\end{figure}

The top panel of Fig.~\ref{fig:embedding} presents the distribution of these pairwise similarities for the two embedding methods. We find that the embeddings by GPT-3 are overly generic with similarities clustering around relatively high values of 0.7--0.9, suggesting a lack of discriminative power (most papers are embedded very similarly). AstroLLaMA's embeddings, on the other hand, exhibit much higher variance within each bin. This suggests that our fine-tuned model is more adept at representing the specialized semantic variance inherent to the field of astronomy, which may enable a more granular representation of astronomical content and can facilitate better document retrieval and semantic analysis.

The bottom panel of Fig.~\ref{fig:embedding} provides two representative examples where AstroLLaMA and GPT-3 classifications diverge. In the first example, GPT-3 fixates on the keyword `magnetized,' resulting in an inflated similarity score, despite the contexts being markedly different. AstroLLaMA, on the other hand, successfully distinguishes between these disparate contexts. In the second example, AstroLLaMA accurately identifies that the study of Spitzer is closely related to star formation. GPT-3, however, fails to make this connection due to the absence of matching keywords.

\section{Limitations and Future Directions}
\label{sec:conclusion}

In this work, we introduce AstroLLaMA, a 7-billion-parameter language model fine-tuned on a dataset encompassing over 300,000 abstracts from astronomical research papers. Compared to its base model, LLaMA-2, and even GPT-4, a current state-of-the-art general LLM, AstroLLaMA exhibits marked improvements in generating high-quality abstracts with a competent grasp of relevant information in this literature.

AstroLLaMA is not without limitations, nevertheless. The most salient is the model's knowledge gaps in certain areas of astronomy: in Fig. \ref{fig:generation_prelim}, AstroLLaMA's estimation of potential star candidates from Gaia-ESO data is notably inaccurate. To address such issues, we are in the process of enriching AstroLLaMA's training set with not just abstracts but the full LaTeX sources of existing astronomy articles, thereby expanding the token count by approximately two orders of magnitude. Another concern lies in the model's tendency to generate hallucinated or fictitious numerical data, an issue likely attributed to our focus on reducing perplexity rather than explicitly steering the model towards factual accuracy. The release of AstroLLaMA aims to facilitate community engagement, both for addressing these inaccuracies and for refining its balance between ``faithfulness'' (respecting scientific evidence and accuracy) and ``creativity'' (being able to come up with interesting hypotheses). 

AstroLLaMA stands as a compelling prototype for specialized LLMs in astronomy, showing superior context-aware capabilities compared to GPT-4 despite having much fewer parameters. It not only paves the way for improved performance in tasks like question-answering, scientific summarization and hypothesis generation but applies also to multi-modal models \citep{liu2023visual}. We have made the AstroLLaMA's weights and its training data publicly available\footnote{\url{https://huggingface.co/universeTBD/astrollama}} for researchers interested in leveraging LLMs for astronomy-centric applications. Along with this, we are establishing various ``playgrounds'' on Hugging Face to invite interested readers to further adapt and refine this robust starting point for a variety of relevant downstream tasks.

\section*{Acknowledgments}

We are deeply grateful to the Microsoft Accelerate Foundation Models Research Initiative for enabling us to fast-track our project. Thanks to advanced AI platform from Microsoft Research, we have been able to significantly expedite our efforts in using language models to analyze astronomical literature.

\newpage
\section*{Ethics Statement}

We obtain the pre-trained weights for LLaMA-2 from Meta, which offers these models for download on Hugging Face. The arXiv dataset used in this paper is publicly available on Kaggle. While we have demonstrated that AstroLLaMA is capable of generating high-quality, relevant abstracts for astronomical research papers, we have noted that it has the potential to generate inaccurate data and measurements. This should serve as a caution for researchers aiming to use this model for downstream tasks, and we invite the adoption of alignment strategies in future work to ameliorate this issue. 

\bibliography{bibtex}
\bibliographystyle{manuscript}

\end{document}